\documentclass[letter,bibyear]{aa} % for the letters

\usepackage{epsfig}
\usepackage{amsmath}
\usepackage{subfigure}
\usepackage{natbib}
\usepackage{multirow}
\usepackage{color}
\usepackage{longtable}
\usepackage{graphicx}
\usepackage{epstopdf}
\usepackage{booktabs}
\usepackage{txfonts}
\newcommand{\jms}{J.~Mol.~Spectrosc.}

\newcommand{\jmst}{J.~Mol.~Struct.}

\newcommand{\kms}{km s$^{-1}$}

\begin{document}

\title{Laboratory and astronomical discovery of cyanothioketene, NCCHCS, in the cold starless core TMC-1\thanks{Based on
observations carried out with the Yebes 40m telescope (projects 19A003, 20A014, 20D023, 21A011, 21D005, and 23A024). The 40m
radio telescope at Yebes Observatory is operated by the Spanish Geographic Institute (IGN; Ministerio de Transportes, Movilidad y Agenda Urbana).}}

\author{
C.~Cabezas\inst{1},
M.~Ag\'undez\inst{1},
Y.~Endo\inst{2},
B.~Tercero\inst{3,4},
Y.-P.~Lee\inst{2,5},
N.~Marcelino\inst{3,4},
P.~de~Vicente\inst{4}, and
J.~Cernicharo\inst{1}
}

\institute{Dept. de Astrof\'isica Molecular, Instituto de F\'isica Fundamental (IFF-CSIC),
C/ Serrano 121, 28006 Madrid, Spain \newline \email carlos.cabezas@csic.es, jose.cernicharo@csic.es
\and Department of Applied Chemistry, Science Building II, National Yang Ming Chiao Tung University, 1001 Ta-Hsueh Rd., Hsinchu 300098, Taiwan
\and Observatorio Astron\'omico Nacional (OAN, IGN), C/ Alfonso XII, 3, 28014, Madrid, Spain
\and Centro de Desarrollos Tecnol\'ogicos, Observatorio de Yebes (IGN), 19141 Yebes, Guadalajara, Spain
\and Center for Emergent Functional Matter Science, National Yang Ming Chiao Tung University, Hsinchu 300093, Taiwan
}

\date{Received; accepted}

\abstract{
We present the detection of cyanothioketene, NCCHCS, in the laboratory and toward TMC-1. This transient species was produced through a discharge of a gas mixture of CH$_2$CHCN and CS$_2$ using argon as carrier gas, and its rotational spectrum between 9 and 40 GHz was characterized using a Balle-Flygare narrowband-type Fourier-transform microwave spectrometer. A total of 21 rotational transitions were detected in the laboratory, all of them exhibiting hyperfine structure induced by the spin of the N nucleus. The spectrum for NCCHCS was predicted in the domain of our line surveys using the derived rotational and distortion constants. The detection in the cold starless core TMC-1 was based on the QUIJOTE$^1$ line survey performed with the Yebes 40m radio telescope. Twenty-three lines were detected with $K_a$=0, 1, and 2 and $J_u$=9 up to 14. The derived column density is (1.2$\pm$0.1)$\times$10$^{11}$ cm$^{-2}$ for a rotational temperature of 8.5$\pm$1\,K. The abundance ratio of thioketene and its cyano derivative, H$_2$CCS/NCCHCS, is 6.5$\pm$1.3. Although ketene is more abundant than thioketene by  $\sim$15 times, its cyano derivative NCCHCO surprisingly is not detected with a 3$\sigma$ upper level to the column density of 3.0$\times$10$^{10}$ cm$^{-2}$, which results in an abundance ratio H$_2$CCO/NCCHCO \,$>$\,430. Hence, the chemistry of CN derivatives seems to be more favored for S-bearing than for O-bearing molecules. We carried out chemical modeling calculations and found that the gas-phase neutral-neutral reactions CCN + H$_2$CS and CN + H$_2$CCS could be a source of NCCHCS in TMC-1.}

\titlerunning{Cyanothioketene in TMC-1}
\authorrunning{Cabezas et al.}

\maketitle

\section{Introduction}
The Taurus molecular cloud, TMC-1, is a cold starless core located in Taurus at a distance of 140 pc \citep{Cernicharo1987}. Its kinetic temperature is $\sim$9\,K \citep{Agundez2023}, and it is known to present a very rich chemistry related to cyanopolyynes
(HC$_{2n+1}$N) and to the radicals C$_n$H and C$_n$N. In addition, aromatic cycles and long hydrocarbons have been unambiguously identified in this cloud \citep[][and references therein]{McGuire2018,McGuire2021,Cernicharo2021a,Cernicharo2021b,Agundez2021,Fuentetaja2022a,Cabezas2022a,
Cernicharo2022a}.

Using the sensitive QUIJOTE\footnote{\textbf{Q}-band \textbf{U}ltrasensitive \textbf{I}nspection \textbf{J}ourney to the \textbf{O}bscure \textbf{T}MC-1 \textbf{E}nvironment} line survey \citep{Cernicharo2021a}, we have reported the discovery of more than 50 new molecular species in this object in the past four years \citep[see, e.g.,][and references therein]{Cernicharo2024}. Most of these molecules are hydrocarbons and their cyano (CN) and ethynyl (CCH) derivatives, together with protonated species of abundant molecules \citep[see, e.g.,][and references therein]{Cabezas2022b,Cabezas2022c,Cernicharo2022b,Cernicharo2024,Agundez2022}.
Last, but not least, about ten of these molecules contain sulphur. Several thio- forms of O-bearing species (formaldehyde, ketene, fulminic acid, isocyanic acid, and cyanic acid), together with other S-bearing cations and radicals, have been also detected in this source \citep{Cernicharo2021c,Cernicharo2021d,Cernicharo2021e,Cabezas2022c,Fuentetaja2022b,Marcelino2023,Cernicharo2024}.
Apparently, not only carbon, but also sulphur therefore has a peculiar chemistry in this cloud, which allows it to produce significant abundances for S-bearing molecular species.

\begin{figure}
\centering
\includegraphics[angle=0,width=0.8\columnwidth]{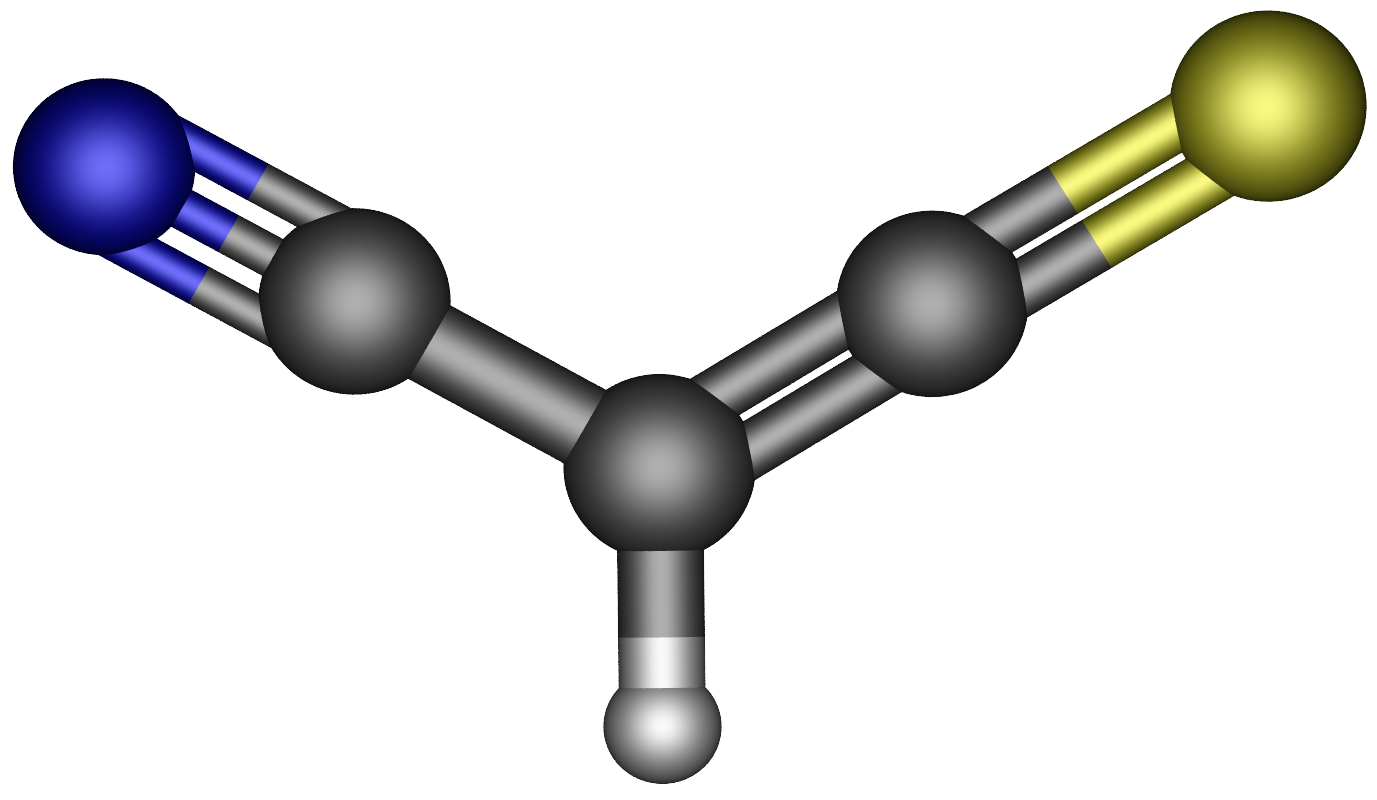}
\caption{Optimized molecular structure of cyanothioketene as obtained from quantum chemical calculations. }
\label{structure}
\end{figure}

The study of CN and CCH derivatives of abundant species can provide insights into the chemical
processes leading to chemical complexity in TMC-1. In this Letter, we report the detection
of the cyano derivative of thioketene (H$_2$CCS). Taking into account the abundance of the latter, its cyano derivative, called cyanothioketene (NCCHCS; see Figure \ref{structure} ), was expected to be present in TMC-1. Based on the observation of the rotational spectrum of cyanothioketene in the laboratory, we could observe a total of 23 lines of this molecule in the QUIJOTE line survey, allowing a solid and robust detection of cyanothioketene in TMC-1. The laboratory experiment and the astronomical observations are described in sections \ref{laboratory} and \ref{astro_obs}, respectively. The detection of the molecule is analyzed in section \ref{astro_detection}, and the chemistry of cyanothioketene in TMC-1 is discussed in section \ref{discussion}.

\section{Laboratory observations}\label{laboratory}

The rotational spectrum of cyanothioketene was observed using a Balle-Flygare narrowband-type narrowband-type Fourier-transform microwave (FTMW) spectrometer operating in the frequency region of 4-40 GHz \citep{Endo1994,Cabezas2016}. The short-lived species NCCHCS was produced in a supersonic expansion by a pulsed electric discharge of a gas mixture of vinyl cyanide (CH$_2$CHCN, 0.3\%) and carbon disulfide (CS$_2$, 0.3\%) diluted in argon. The gas mixture was flowed through a pulsed-solenoid valve that is accommodated in the backside of one of the cavity mirrors and aligned parallel to the optical axis of the resonator. A pulse voltage of 2000 V with a duration of 450 $\mu$s was applied between stainless-steel electrodes attached to the exit of the pulsed discharge nozzle, resulting in an electric discharge synchronized with the gas expansion. The resulting products generated in the discharge were then probed by FTMW spectroscopy capable of resolving small hyperfine splittings.

Prior to the experimental work, we performed quantum chemical calculations to obtain the theoretical spectroscopic parameters of cyanothioketene, which were used to guide the spectral search. We used the coupled cluster method CCSD(T)-F12/cc-pCVTZ-F12 with all electrons (valence and core) correlated to derive the rotational constants and the dipole moment components. These calculations were performed using the MOLPRO program \citep{Werner2020}. On the other hand, the $^{14}$N hyperfine constants were calculated using the MP2/cc-pVTZ level of theory, and the vibration-rotation interaction constants were calculated using the B3LYP/cc-pVTZ level of theory. All these calculations were carried out using the Gaussian16 program package \citep{Frisch2016}. The results are shown in Table \ref{theory}.

\begin{table}
\small
\centering
\caption{Predicted molecular constants (all in MHz) of NCCHCS.}
\label{theory}
\centering
\begin{tabular}{{lcc}}
\hline
\hline
Constant           & Optimization & Opt+Vib-rot \\
\hline
$A$                &  23095     &      23613     \\
$B$                &  1799      &      1790      \\
$C$                &  1669      &      1661      \\
$\chi_{aa}$        &  -1.87     &                \\
$\chi_{bb}$        &   0.29     &                \\
\hline
\hline
\end{tabular}
%\tablefoot{\\
%\tablefoottext{a}{The uncertainties (in parentheses) are in units of the last
%significant digits.}\\
%}
\end{table}
\normalsize

\begin{figure}
\centering
\includegraphics[angle=0,width=0.48\textwidth]{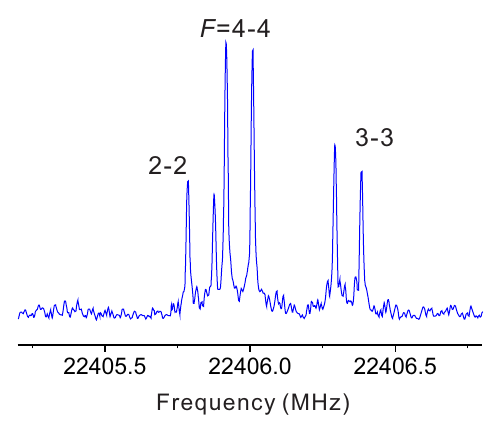}
\caption{FTMW hyperfine spectrum of 3$_{1,2}$–3$_{0,3}$ rotational transition for NCCHCS. Each hyperfine component is split into two Doppler components because the direction of the supersonic jet expansion is parallel to the standing wave in the Fabry-P\'erot cavity of the spectrometer. The spectra were achieved by 300 shots of accumulation and a step scan of 1.0 MHz with a repetition rate of 10 Hz.} \label{spectrum}
\end{figure}

We first searched for the $a$-type rotational transitions since the calculated dipole moment component $\mu_a$ is higher than $\mu_b$, 3.38 versus 1.95\,D. Rotational transitions with $K_a$= 0 and 1 were found close to the frequency predictions and easily assigned because the hyperfine pattern due to the nuclear quadrupole coupling effects was very similar to that predicted by our calculations. A total of 15 $a$-type rotational transitions with $K_a$= 0 and 1 were observed in the 10-24 GHz region. The low temperature, $\sim$2.5 K, of our supersonic expansion means that the energy levels for the lines with $K_a$= 2 and higher are not enough populated, preventing their observation. In addition to the $a$-type transitions, we measured six $b$-type lines (see Figure \ref{spectrum}). Therefore, the final dataset consisted of 88 hyperfine components corresponding to 21 pure rotational transitions. Table \ref{tab_lab_ncchcs} contains the experimental frequencies for all the observed hyperfine components.  Rotational, centrifugal, and nuclear quadrupole coupling constants were determined by fitting the transition frequencies with the SPFIT program \citep{Pickett1991} to a Watson's $S$-reduced Hamiltonian for asymmetric top molecules, with the following form: $H$ = $H_R$ + $H_Q$ \citep{Watson1977}, where $H_R$ contains the rotational and centrifugal distortion parameters, and $H_Q$ contains the quadrupole coupling interactions. The energy levels involved in each transition are labeled with the quantum numbers $J$, $K_{a}$, $K_{c}$, and $F$, where \textbf{F} = \textbf{J} + \textbf{I}(N), and \textbf{I}(N) = 1. The analysis rendered the experimental constants listed in Table \ref{rotationalconstants}.

\begin{table}
\tiny
\centering
\caption{Molecular constants (all in MHz) of NCCHCS.}
\label{rotationalconstants}
\centering
\begin{tabular}{{lccc}}
\hline
\hline
Constant$^a$         & Laboratory        & TMC-1         & Merged Fit \\
\hline
$A_0$                & 23743.58026(105)  & 23744.47(74)  & 23743.58064(90) \\
$B_0$                & 1785.880959(232)  & 1785.88036(96)& 1785.880880(115)\\
$C_0$                & 1658.684211(190)  & 1658.6848(10) & 1658.684441(98) \\
10$^4$\,$D_J$        & 6.5243(142)       & 6.544(11)     & 6.5385(37)      \\
10$^2$\,$D_{JK}$     &-6.4936(52)        &-6.5052(75)    & -6.4959(34)     \\
10$^4$\,$d_1$        & 1.3199(127)       & 1.301(15)     & 1.3088(42)      \\
$\chi_{aa}$          & -1.9970(100)      &               & -1.9965(98)     \\
$\chi_{bb}$          &  0.3827(72)       &               & 0.3823(70)      \\
$N_{lines}$          & 65                & 23            & 88              \\
$\sigma$(kHz)        & 2.4               &  8.4          & 4.5             \\
$\nu_{max}$(GHz)     & 25.4              & 49.0          & 49.0            \\
\hline
\hline
\end{tabular}
\tablefoot{\\
\tablefoottext{a}{The uncertainties (in parentheses) are in units of the last
significant digits.}\\
}
\end{table}
\normalsize

\section{Astronomical observations}\label{astro_obs}

The observational data used in this work are part of QUIJOTE \citep{Cernicharo2021a}, a spectral line survey of TMC-1 in the Q band carried out with the Yebes 40m telescope at the position $\alpha_{J2000}=4^{\rm h} 41^{\rm  m} 41.9^{\rm s}$ and $\delta_{J2000}= +25^\circ 41' 27.0''$, corresponding to the cyanopolyyne peak (CP) in TMC-1. The receiver was built within the Nanocosmos project\footnote{\texttt{https://nanocosmos.iff.csic.es/}} and consists of two cold high-electron mobility transistor amplifiers covering the 31.0-50.3 GHz band with horizontal and vertical polarizations. The receiver temperatures achieved in the 2019 and 2020 runs varied from 22 K at 32 GHz to 42 K at 50 GHz. Some power adaptation in the down-conversion chains reduced the receiver temperatures during 2021 to 16\,K at 32 GHz and 30\,K at 50 GHz. The backends were $2\times8\times2.5$ GHz fast Fourier transform spectrometers with a spectral resolution of 38 kHz, providing the whole coverage of the Q band in both polarizations.  A more detailed description of the system is given by \citet{Tercero2021}.

The data of the QUIJOTE line survey presented here were gathered in several observing runs between November 2019 and July 2023. All observations were performed using a frequency-switching observing mode with a frequency throw of 10 and 8 MHz. The total observing time on the source for data taken with frequency throws of 10 MHz and 8 MHz was 465 and 737 hours, respectively. Hence, the total observing time on source was 1202 hours. The measured sensitivity varied between 0.08 mK at 32 GHz and 0.2 mK at 49.5 GHz. The sensitivity of QUIJOTE is about 50 times better than that of previous line surveys in the Q band of TMC-1 \citep{Kaifu2004}. For each frequency throw, different local oscillator frequencies were used in order to remove possible sideband effects in the down-conversion chain. A detailed description of the QUIJOTE line survey is provided in \citet{Cernicharo2021a}. The data analysis procedure has been described by \citet{Cernicharo2022a}.

The main-beam efficiency measured during our observations in 2022 varied from 0.66 at 32.4 GHz to 0.50 at 48.4 GHz \citep{Tercero2021} and can be given across the Q-Band by $B_{\rm eff}$=0.797 exp[$-$($\nu$(GHz)/71.1)$^2$]. The forward telescope efficiency is 0.97. The telescope beam size at half-power intensity is 54.4$''$ at 32.4 GHz and 36.4$''$ at 48.4 GHz. The intensity scale used in this study is the antenna temperature ($T_A^*$). Calibration was performed using two absorbers at different temperatures and the atmospheric transmission model ATM \citep{Cernicharo1985, Pardo2001}. The absolute calibration uncertainty is 10\,$\%$. However, the relative calibration between lines within the QUIJOTE survey is probably better because all the spectral features have common pointing and calibration errors. The data were analyzed with the GILDAS package\footnote{\texttt{http://www.iram.fr/IRAMFR/GILDAS}}.

\section{Astronomical detection of NCCHCS in TMC-1}\label{astro_detection}

The line identification was performed using the MADEX code \citep{Cernicharo2012}, the CDMS catalog \citep{Muller2005}, and the JPL catalog \citep{Pickett1998}.

\begin{figure*}[h]
\centering
\includegraphics[width=\textwidth]{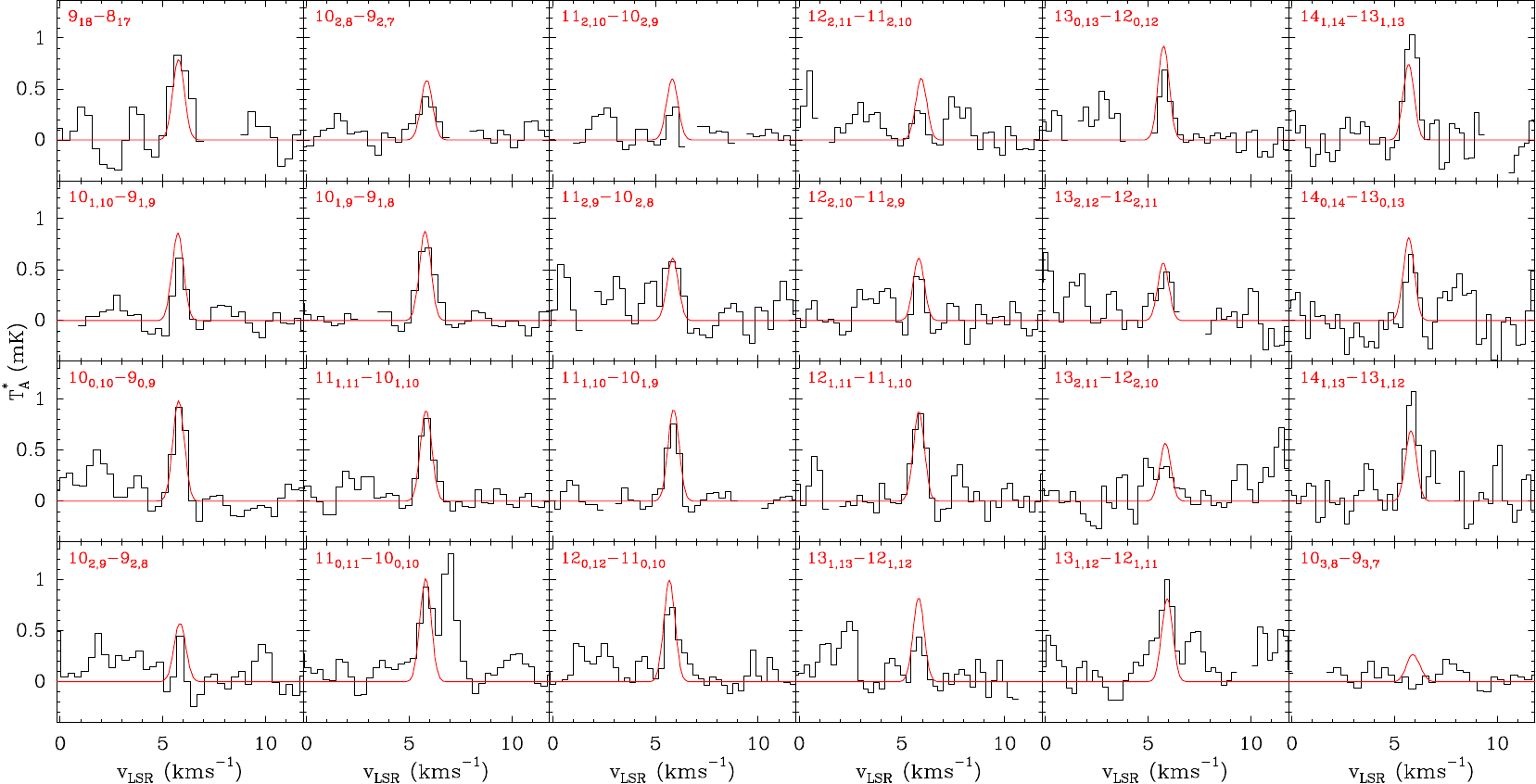}
\caption{Observed lines of NCCHCS in TMC-1. The abscissa corresponds to the velocity
scale that was derived using the frequencies given in Table \ref{line_parameters} for which a v$_{LSR}$ of 5.83
kms$^{-1}$ was adopted \citep{Cernicharo2020}. The ordinate is the antenna temperature in milli Kelvin. The blank channels correspond to negative features produced in the folding of the frequency-switching data. The red spectra correspond to the synthetic line profiles modeled with a rotational temperature of 8.5\,K and a column density
of 1.2$\times$10$^{11}$ cm$^{-2}$.}
\label{ncchcs_tmc1}
\end{figure*}

Using the frequencies predicted from the laboratory data (see Sect. \ref{laboratory} and Table \ref{rotationalconstants}), we searched for the $K_a$=0 and 1 lines in our QUIJOTE line survey. Sixteen lines should be within our data with $J$=9 up to 14. Fifteen of them were easily detected in the line survey, and one line is fully blended with a strong line from H$_2$CCCS and can therefore not be measured. After we were fully confident of the presence of NCCHCS in TMC-1, we searched for the $K_a$=2 lines and detected eight of them with $J$=10, 11, 12, and 13. Consequently, 23 lines of NCCHCS were observed with the QUIJOTE line survey, providing a unambiguous detection of this species in space. The line intensities range from 0.4 mK to $\sim$1.1 mK, and they are shown in Fig. \ref{ncchcs_tmc1}. The derived line parameters are given in Table \ref{line_parameters}. The hyperfine structure is not resolved for any of the observed lines. Nevertheless, the lines are slightly broadened by $\sim$15-20 kHz. We also searched for $K_a$=3 lines, but their predicted intensities are below the sensitivity of our data (see the bottom right panel of Fig. \ref{ncchcs_tmc1}).

We fit the frequencies measured in TMC-1 using the standard Watson Hamiltonian in the $S$-representation. The resulting rotational and distortion constants are given in Table \ref{rotationalconstants}. They agree excellently with those derived in the laboratory (see the first column of the same table). A merged fit to the laboratory and TMC-1 data provides improved values for these parameters. In this fit, we consider uncertainties of 3 kHz for the laboratory lines, and for the transitions observed in TMC-1, the uncertainties range between 10 and 20 kHz (see Table \ref{line_parameters}). The resulting values are given in the fourth column of Table \ref{rotationalconstants}. These are the recommended constants for computing the rotational spectrum of cyanothioketene. We considered that frequency predictions could be acceptable for radio astronomical purposes up to 100 GHz and $K_a\le$3. The catalog file with the predicted frequencies and the calculated intensities up to $J$ = 30 at 300\,K is provided in Table A.2 at the CDS.

In order to estimate the column densities and rotational temperatures, we assumed that the source has a diameter of 80$''$ \citep{Fosse2001} and that it has an uniform brightness distribution for all observed lines. The adopted intrinsic line width is 0.6 km\,s$^{-1}$. The best fit to all the observed intensities and line profiles, which includes the hyperfine structure of the lines to take into account the small line broadening, provides a rotational temperature of 8.5$\pm$1\,K and a column density for NCCHCS of (1.2$\pm$0.1)$\times$10$^{11}$ cm$^{-2}$. The total column density of thioketene in TMC-1 is  (7.8$\pm$0.8)$\times$10$^{11}$ cm$^{-2}$ \citep{Cernicharo2021d}. Hence, the abundance ratio of H$_2$CCS and its CN derivative NCCHCS is 6.5$\pm$1.3. This ratio is lower by a factor of $\sim$5 than the equivalent ratio for thioformaldehyde, H$_2$CS/HCSCN, $\sim$35 \citep{Cernicharo2021d}. The derived column densities and those of related species are given in Table \ref{columndensities}.

\begin{table}
\centering
\caption{Column densities of H$_2$C$_n$O and H$_2$C$_n$S and their CN and CCH derivatives}
\label{columndensities}
\centering
\begin{tabular}{{lccc}}
\hline
[Y]$^a$&         N[Y]&           N[(Y-H)+CN]$^b$  &    N[(Y-H)+CCH]$^c$\\
\hline
H$_2$CO     &  5.0$\times$10$^{14}$ (0) &     3.5$\times$10$^{11}$ (1)    &    1.5$\times$10$^{12}$ (1)\\
H$_2$CS     &  3.7$\times$10$^{13}$ (1) &     1.3$\times$10$^{12}$ (1)    &    3.2$\times$10$^{11}$ (1)\\
H$_2$CCO    &  1.3$\times$10$^{13}$ (2) &    $\le$3.0$\times$10$^{10}$ (3)&    (4)                     \\
H$_2$CCS    &  7.8$\times$10$^{11}$ (5) &     1.2$\times$10$^{11}$ (3)    &    (4)                     \\
HCO         &  1.1$\times$10$^{12}$ (6) &    $\le$3$\times$10$^{10}$ (3)  &  1.3$\times$10$^{11}$ (6)     \\
HCS         &  5.5$\times$10$^{12}$ (5) &    $\le$3$\times$10$^{10}$ (3)  &  $\le$2.4$\times$10$^{11}$ (5) \\               &                    \\
%CH2CN+HCO/HCS; CH2N+HCO/HCS
\hline
\end{tabular}
\tablefoot{\\
\tablefoottext{a}{Molecular species denoted as [Y] (Y being HCO, HCS, H$_2$CO, H$_2$CS, H$_2$CCO or H$_2$CCS.}\\
\tablefoottext{b}{Column density in cm$^{-2}$ when one hydrogen of CH$_2$ (or CH) is replaced by CN.
Reference for this value is between parentheses.}\\
\tablefoottext{c}{Column density in cm$^{-2}$ when one hydrogen of CH$_2$ (or CH) is replaced by CCH.
Reference for this value is between parentheses.}\\
\tablefoottext{0}{\citet{Agundez2013}}
\tablefoottext{1}{\citet{Cernicharo2021e}}
\tablefoottext{2}{\citet{Soma2018}}
\tablefoottext{3}{This work.}
\tablefoottext{4}{Low dipole moment species. This work.}
\tablefoottext{5}{\citet{Cernicharo2021d}}
\tablefoottext{6}{\citet{Cernicharo2021f}}
}
\end{table}

Like for the cyano and ethynyl derivatives of H$_2$CS, HCSCN and HCSCCH \citep{Cernicharo2021e}, we also expect
a significant abundance for the CCH derivative of thioketene, HCCCHCS. However, our quantum chemical calculations, carried out at the same level of theory as those for NCCHCS, indicate that for HCCCHCS, the $\mu_a$ and $\mu_b$ dipole moment components are very low, 0.1 and 0.3\,D, respectively. Hence, the lines for HCCCHCS are expected to be below our detection limit, and no attempt was made to observe this species in the laboratory.

\section{Discussion}\label{discussion}

It might be thought that if NCCHCS is produced in TMC-1 from the reaction of CN and H$_2$CCS, then we could expect to have NCCHCO as well, which could be obtained from the reaction of CN with H$_2$CCO. Rotational spectroscopy for NCCHCO is available from \citet{Hahn2004} and \citet{Margules2020}. The molecule was searched for toward several interstellar sources, including TMC-1, but without success \citep{Margules2020}. Their 3$\sigma$ upper limit to the column density of NCCHCO in TMC-1 is 3.3$\times$10$^{11}$ cm$^{-2}$. Our more sensitive data in TMC-1 do not produce a positive detection of this species. For a line width of 0.6 km\,s$^{-1}$ and an excitation temperature similar to that of NCCHCS, the derived 3$\sigma$ upper limit from the six individual lines of NCCHCO with the best chances to be detected ($J_u$=6 and 7, $K_a$=0, 1) is  N(NCCHCO)$\le$3$\times$10$^{10}$ cm$^{-2}$. Hence, the abundance ratio NCCHCS/NCCHCO is $\ge$4. With the column density derived by \citet{Soma2018} for ketene (see Table \ref{columndensities}), the H$_2$CCO/NCCHCO abundance ratio is $\ge$430, which contrasts with the much lower value of 6.5 found for H$_2$CCS/NCCHCS.

Alternative paths to form NCCHCS could involve the species HCCS, CCS, H$_2$CS, HCS, and CS. It is worth noting that the radical CCS of these species is much more abundant than its oxygenated counterpart CCO \citep[CCS/CCO\,$\sim$\,73,][]{Cernicharo2021d,Cernicharo2021f}. The radical HCS is also more abundant in TMC-1 than HCO, by a factor 5 (see Table \ref{columndensities}). Therefore, if either CCS or HCS act as a precursor of NCCHCS, the higher abundance of the sulfur species compared to the oxygenated counterpart could explain why NCCHCS is much more abundant than NCCHCO in TMC-1 in the scenario in which the reaction rate coefficients for the sulfur and oxygen reactions are similar. Alternatively, the sulfur and oxygen reactions could behave differently, favoring the formation of NCCHCS, but not that of NCCHCO.

\begin{figure}
\centering
\includegraphics[angle=0,width=\columnwidth]{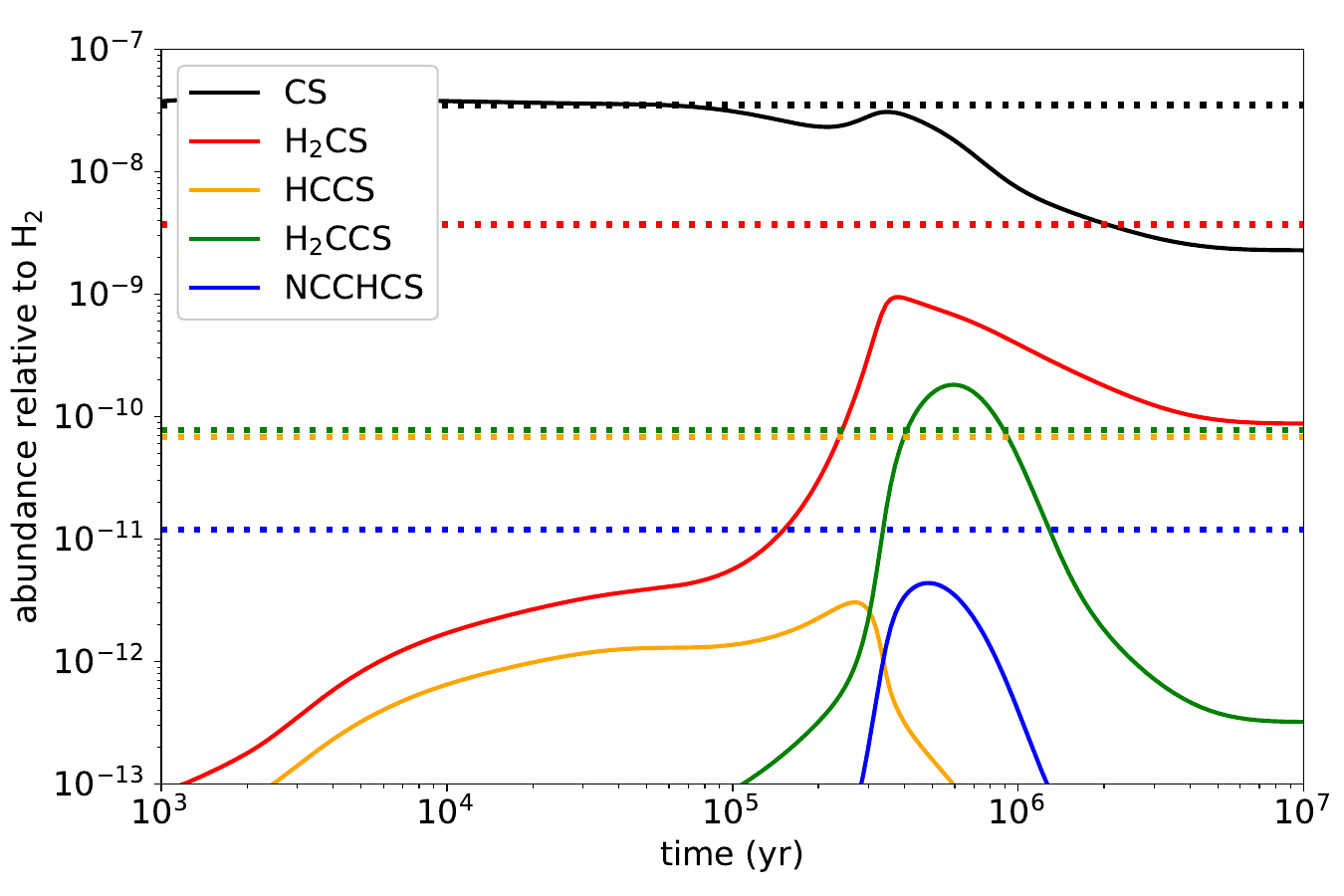}
\caption{Calculated fractional abundances for NCCHCS and possible S-bearing precursors as a function of time. The horizontal dotted lines correspond to observed values in TMC-1 adopting a column density of H$_2$ of 10$^{22}$ cm$^{-2}$ \citep{Cernicharo1987}.}
\label{fig:abun}
\end{figure}

In order to shed additional light on the formation of cyanothioketene in TMC-1, we carried out chemical modeling calculations adopting as starting point the model developed in \cite{Cernicharo2021d} to explain the detection of H$_2$CCS in TMC-1. The core of the chemical network was taken from the UMIST database \citep{McElroy2013} and \cite{Vidal2017}. Here, we considered several neutral-neutral reactions occurring with H atom elimination to evaluate their potential role in the formation of NCCHCS. These reactions are CN + H$_2$CCS, H$_2$CN + CCS,  CCN + H$_2$CS, and HCCN + HCS. All these reactions are exothermic when producing NCCHCS + H. We note that the reactions HCN + HCCS and CH$_2$CN + CS were calculated to be endothermic by 21.3 and 11.5 kcal mol$^{-1}$, respectively, when forming NCCHCS + H, and were thus not considered. We assumed that the four exothermic neutral-neutral reactions considered have no barriers, and thus, we adopted a rate coefficient of 4\,$\times$\,10$^{-10}$ cm$^3$ s$^{-1}$, which is typically found for reactions without barriers at low temperatures (e.g., \citealt{Sims1993}). In addition, we assumed that NCCHCS is mainly destroyed through reactions with C atoms and abundant cations, such as HCO$^+$ and H$_3^+$. The results from the chemical model are shown in Fig.\,\ref{fig:abun}. The peak abundance calculated for NCCHCS agrees very well with the observed value, although this may be fortuitous, considering the large uncertainties associated with the chemistry of this molecule.

According to the chemical model, the reactions H$_2$CN + CCS and HCCN + HCS contribute little to the formation of NCCHCS, and thus the main pathways are CCN + H$_2$CS and CN + H$_2$CCS. To our knowledge, none of these two reactions have been studied theoretically or experimentally, and thus it is difficult to anticipate which might be the main route to cyanothioketene in TMC-1. Of the analogous reactions involving oxygen rather than sulfur, the only reaction that has been studied to some extent is CN + H$_2$CCO. Measurements of the rate coefficient over the temperature range 296-567 K \citep{Edwards1998} yielded a negative temperature dependence, which suggests that it is likely fast at low temperatures. However, quantum chemical calculations \citep{Sun2006,Zhang2006,Zhao2008} indicate that the main products are CH$_2$CN + CO, rather than NCCHCO + H. If the reaction CN + H$_2$CCS behaves similarly and the main products are CH$_2$CN + CS, it would not be a source of NCCHCS in TMC-1. In summary, our analysis indicates that the reactions CCN + H$_2$CS and CN + H$_2$CCS could lead to cyanothioketene. Studies of the kinetics of these reactions are highly desirable to shed light on the formation of NCCHCS in the cold interstellar medium. Finally, we note that the reaction HCCN + HCO has been studied theoretically by \cite{Ballotta2023}, who found that it can lead without a barrier to cyanoketene (NCCHCO). Our chemical model predicts that the analogous reaction involving sulfur, HCCN + HCS, even if it is fast and leads to NCCHCS, should be a minor route to cyanothioketene in TMC-1, however.

\section{Conclusions}

We reported the discovery of the NCCHCS molecule in TMC-1. This a cyano derivative of thioketene was detected based on the accurate results from a laboratory rotational spectroscopic study. A total of 21 rotational transitions were measured in the laboratory, which allowed us to accurately characterize this transient species. Consequently, 23 lines of NCCHCS were observed with the QUIJOTE line survey, providing an unambiguous detection of this species in space, with a derived column density of (1.2$\pm$0.1)$\times$10$^{11}$ cm$^{-2}$ for a rotational temperature of 8.5$\pm$1\,K. Although ketene is much more abundant than thioketene, its cyano derivative is not detected in TMC-1, which indicates that the chemistry of CN derivatives seems to be more favored for S-bearing than for O-bearing molecules. Our chemical modeling calculations indicate that the gas-phase neutral-neutral reactions CCN + H$_2$CS and CN + H$_2$CCS could be a source of NCCHCS in TMC-1.

\begin{acknowledgements}

The present study was supported by Ministry of Science and Technology of Taiwan and Consejo Superior de Investigaciones Cient\'ificas for funding support under the MoST-CSIC Mobility Action 2021 (Grant 11-2927-I-A49-502 and OSTW200006). C.C., M.A, and J.C. thank Ministerio de Ciencia e Innovaci\'on of Spain (MICIU) for funding support through projects PID2019-106110GB-I00, PID2019-107115GB-C21 / AEI / 10.13039/501100011033, and PID2019-106235GB-I00. C.C., M.A, and J.C. also thank ERC for funding through grant ERC-2013-Syg-610256-NANOCOSMOS. Y. E. acknowledges Ministry of Science and Technology of Taiwan for MOST 104-2113-M-009-202 project.

\end{acknowledgements}

\normalsize
\begin{appendix}
\onecolumn

\section{Laboratory-observed frequencies}

The laboratory experiments described in Sect. \ref{laboratory} have permitted us to measure 88 hyperfine components, corresponding to 21 purely rotational transitions for the NCCHCS molecule. The observed frequencies, the differences between observed and calculated frequencies, and the quantum number assignments are given in Table \ref{tab_lab_ncchcs}.

\begin{table}
\small
\caption{Laboratory-observed transition frequencies for NCCHCS.}
\label{tab_lab_ncchcs}
\centering
\begin{tabular}{cccccccccr}
\hline
\hline
 $J'$ & $K'_a$ & $K'_c$ & $F'$  & $J''$ & $K''_a$ & $K''_c$ & $F''$ & $\nu_{obs}$  &  Obs-Calc \\
   &   &        &      &        &        &                  &    &   (MHz)      &   (MHz)      \\
\hline
3	&	1	&	3	&	3	&	2	&	1	&	2	&	2	&	10142.765 &   0.000  \\
3	&	1	&	3	&	2	&	2	&	1	&	2	&	1	&	10142.914 &   0.004  \\
3	&	1	&	3	&	4	&	2	&	1	&	2	&	3	&	10142.952 &   0.001  \\
3	&	0	&	3	&	2	&	2	&	0	&	2	&	1	&	10331.323 &   0.001  \\
3	&	0	&	3	&	3	&	2	&	0	&	2	&	2	&	10331.422 &   0.001  \\
3	&	0	&	3	&	4	&	2	&	0	&	2	&	3	&	10331.447 &   0.001  \\
3	&	1	&	2	&	3	&	2	&	1	&	1	&	2	&	10524.325 &   0.000  \\
3	&	1	&	2	&	4	&	2	&	1	&	1	&	3	&	10524.498 &   0.001  \\
3	&	1	&	2	&	2	&	2	&	1	&	1	&	1	&	10524.530 &  -0.002  \\
4	&	1	&	4	&	4	&	3	&	1	&	3	&	3	&	13523.108 &   0.002  \\
4	&	1	&	4	&	3	&	3	&	1	&	3	&	2	&	13523.146 &  -0.001  \\
4	&	1	&	4	&	5	&	3	&	1	&	3	&	4	&	13523.190 &  -0.001  \\
4	&	0	&	4	&	3	&	3	&	0	&	3	&	2	&	13772.547 &   0.003  \\
4	&	0	&	4	&	4	&	3	&	0	&	3	&	3	&	13772.582 &  -0.002  \\
4	&	0	&	4	&	5	&	3	&	0	&	3	&	4	&	13772.603 &   0.002  \\
4	&	1	&	3	&	4	&	3	&	1	&	2	&	3	&	14031.816 &   0.000  \\
4	&	1	&	3	&	3	&	3	&	1	&	2	&	2	&	14031.883 &  -0.001  \\
4	&	1	&	3	&	5	&	3	&	1	&	2	&	4	&	14031.894 &   0.002  \\
5	&	1	&	5	&	5	&	4	&	1	&	4	&	4	&	16902.802 &  -0.002  \\
5	&	1	&	5	&	4	&	4	&	1	&	4	&	3	&	16902.822 &   0.004  \\
5	&	1	&	5	&	6	&	4	&	1	&	4	&	5	&	16902.850 &  -0.001  \\
5	&	0	&	5	&	4	&	4	&	0	&	4	&	3	&	17211.460 &  -0.005  \\
5	&	0	&	5	&	5	&	4	&	0	&	4	&	4	&	17211.485 &   0.000  \\
5	&	0	&	5	&	6	&	4	&	0	&	4	&	5	&	17211.501 &   0.003  \\
5	&	1	&	4	&	5	&	4	&	1	&	3	&	4	&	17538.625 &   0.001  \\
5	&	1	&	4	&	4	&	4	&	1	&	3	&	3	&	17538.649 &  -0.005  \\
5	&	1	&	4	&	6	&	4	&	1	&	3	&	5	&	17538.668 &   0.003  \\
6	&	1	&	6	&	6	&	5	&	1	&	5	&	5	&	20281.750 &  -0.002  \\
6	&	1	&	6	&	5	&	5	&	1	&	5	&	4	&	20281.750 &  -0.008  \\
6	&	1	&	6	&	7	&	5	&	1	&	5	&	6	&	20281.783 &   0.002  \\
6	&	0	&	6	&	5	&	5	&	0	&	5	&	4	&	20647.545 &  -0.006  \\
6	&	0	&	6	&	6	&	5	&	0	&	5	&	5	&	20647.556 &  -0.006  \\
6	&	0	&	6	&	7	&	5	&	0	&	5	&	6	&	20647.577 &   0.003  \\
6	&	1	&	5	&	6	&	5	&	1	&	4	&	5	&	21044.615 &  -0.005  \\
6	&	1	&	5	&	5	&	5	&	1	&	4	&	4	&	21044.637 &   0.001  \\
6	&	1	&	5	&	7	&	5	&	1	&	4	&	6	&	21044.648 &   0.002  \\
1	&	1	&	0	&	2	&	1	&	0	&	1	&	2	&	22084.848 &   0.003  \\
2	&	1	&	1	&	1	&	2	&	0	&	2	&	1	&	22212.625 &   0.001  \\
2	&	1	&	1	&	3	&	2	&	0	&	2	&	3	&	22212.914 &   0.001  \\
2	&	1	&	1	&	2	&	2	&	0	&	2	&	2	&	22213.436 &   0.003  \\
3	&	1	&	2	&	2	&	3	&	0	&	3	&	2	&	22405.832 &  -0.001  \\
3	&	1	&	2	&	4	&	3	&	0	&	3	&	4	&	22405.964 &   0.000  \\
3	&	1	&	2	&	3	&	3	&	0	&	3	&	3	&	22406.339 &   0.000  \\
4	&	1	&	3	&	3	&	4	&	0	&	4	&	3	&	22665.173 &   0.000  \\
4	&	1	&	3	&	5	&	4	&	0	&	4	&	5	&	22665.254 &  -0.001  \\
4	&	1	&	3	&	4	&	4	&	0	&	4	&	4	&	22665.571 &   0.000  \\
5	&	1	&	4	&	4	&	5	&	0	&	5	&	4	&	22992.363 &   0.001  \\
5	&	1	&	4	&	6	&	5	&	0	&	5	&	6	&	22992.421 &   0.000  \\
5	&	1	&	4	&	5	&	5	&	0	&	5	&	5	&	22992.711 &   0.001  \\
6	&	1	&	5	&	5	&	6	&	0	&	6	&	5	&	23389.448 &   0.001  \\
6	&	1	&	5	&	7	&	6	&	0	&	6	&	7	&	23389.492 &  -0.001  \\
6	&	1	&	5	&	6	&	6	&	0	&	6	&	6	&	23389.767 &  -0.001  \\
7	&	1	&	7	&	7	&	6	&	1	&	6	&	6	&	23659.820 &  -0.001  \\
7	&	1	&	7	&	6	&	6	&	1	&	6	&	5	&	23659.820 &  -0.003  \\
7	&	1	&	7	&	8	&	6	&	1	&	6	&	7	&	23659.844 &   0.002  \\
7	&	1	&	6	&	8	&	7	&	0	&	7	&	8	&	23858.889 &   0.002  \\
7	&	1	&	6	&	7	&	7	&	0	&	7	&	7	&	23859.154 &  -0.002  \\
7	&	0	&	7	&	6	&	6	&	0	&	6	&	5	&	24080.259 &   0.005  \\
7	&	0	&	7	&	7	&	6	&	0	&	6	&	6	&	24080.261 &   0.000  \\
7	&	0	&	7	&	8	&	6	&	0	&	6	&	7	&	24080.272 &   0.001  \\
7	&	1	&	6	&	7	&	6	&	1	&	5	&	6	&	24549.646 &  -0.001  \\
7	&	1	&	6	&	6	&	6	&	1	&	5	&	5	&	24549.656 &   0.000  \\
7	&	1	&	6	&	8	&	6	&	1	&	5	&	7	&	24549.667 &   0.003  \\
1	&	1	&	1	&	2	&	0	&	0	&	0	&	1	&	25402.370 &  -0.004  \\
1	&	1	&	1	&	1	&	0	&	0	&	0	&	1	&	25402.485 &  -0.003  \\
\hline
\hline
\end{tabular}
\end{table}
\normalsize

\section{Line parameters}
The line parameters for all observed transitions were derived by fitting a Gaussian line profile to them
using the GILDAS package. A velocity range of $\pm$20\,\kms\, around each feature was considered for the fit after a polynomial baseline was removed. Negative features produced in the folding of the frequency switching data were blanked
before baseline removal. The derived line parameters for TMC-1 are given in Table \ref{line_parameters}.

\begin{table*}[h]
\centering
\caption{Line parameters for all observed transitions for NCCHCS.}
\label{line_parameters}
\begin{tabular}{lcccccccc}
\hline
Transition            & $\nu_{obs}$~$^a$ & $\int T_A^* dv$~$^b$ & $\Delta v$~$^c$ & $T_A^*$~$^d$ \\
                      & (MHz)              & (mK\,km\,s$^{-1}$)  & (km\,s$^{-1}$)  & (mK) \\
\hline
$ 9_{1, 8}-8_{1,7}$  &31556.110$\pm$0.010& 0.92$\pm$0.15&  1.00$\pm$0.17& 0.86$\pm$0.16&A\\
$10_{1,10}-9_{1,9}$  &33787.451$\pm$0.010& 0.40$\pm$0.06&  0.59$\pm$0.09& 0.63$\pm$0.09&\\
$10_{0,10}-9_{0,9}$  &34352.683$\pm$0.010& 0.73$\pm$0.07&  0.70$\pm$0.07& 0.98$\pm$0.09&\\
$10_{2,9}-9_{2,8}$   &34434.336$\pm$0.020& 0.15$\pm$0.07&  0.33$\pm$0.10& 0.43$\pm$0.15&A\\
$10_{2,8}-9_{2,7}$   &34524.679$\pm$0.010& 0.49$\pm$0.11&  1.17$\pm$0.31& 0.39$\pm$0.09&\\
$10_{1,9}-9_{1,8}$   &35057.192$\pm$0.010& 0.76$\pm$0.10&  0.92$\pm$0.13& 0.77$\pm$0.07&\\
$11_{1,11}-10_{1,10}$&37160.746$\pm$0.010& 0.70$\pm$0.08&  0.80$\pm$0.11& 0.82$\pm$0.08&\\
$11_{0,11}-10_{0,10}$&37766.565$\pm$0.010& 0.70$\pm$0.08&  0.89$\pm$0.15& 0.90$\pm$0.08&\\
$11_{2,10}-10_{2,9}$ &37873.642$\pm$0.020& 0.18$\pm$0.03&  0.30$\pm$0.12& 0.57$\pm$0.06&\\
$11_{2,9}-10_{2,8}$  &37993.801$\pm$0.020& 0.59$\pm$0.12&  0.85$\pm$0.18& 0.65$\pm$0.15&A\\
$11_{1,10}-10_{1,9}$ &38556.602$\pm$0.010& 0.57$\pm$0.06&  0.67$\pm$0.07& 0.79$\pm$0.08&\\
$12_{1,12}-11_{1,11}$&                   &              &               &              &B\\
$12_{0,12}-11_{0,11}$&41174.511$\pm$0.010& 0.67$\pm$0.08&  0.88$\pm$0.14& 0.71$\pm$0.07&\\
$12_{2,11}-11_{2,10}$&41311.794$\pm$0.020& 0.23$\pm$0.05&  0.66$\pm$0.15& 0.33$\pm$0.09&\\
$12_{2,10}-11_{2,9}$ &41467.488$\pm$0.020& 0.27$\pm$0.06&  0.46$\pm$0.14& 0.54$\pm$0.11&\\
$12_{1,11}-11_{1,10}$&42054.115$\pm$0.010& 0.70$\pm$0.10&  0.74$\pm$0.11& 0.89$\pm$0.10&C\\
$13_{1,13}-12_{1,12}$&43902.762$\pm$0.010& 0.28$\pm$0.06&  0.57$\pm$0.12& 0.46$\pm$0.12&C\\
$13_{0,13}-12_{0,12}$&44576.159$\pm$0.010& 0.47$\pm$0.10&  0.67$\pm$0.11& 0.65$\pm$0.09&\\
$13_{2,12}-12_{2,11}$&44748.620$\pm$0.020& 0.30$\pm$0.09&  0.63$\pm$0.21& 0.46$\pm$0.15&\\
$13_{2,11}-12_{2,10}$&44946.028$\pm$0.020& 0.38$\pm$0.10&  1.08$\pm$0.33& 0.33$\pm$0.11&\\
$13_{1,12}-12_{1,11}$&45549.568$\pm$0.010& 1.04$\pm$0.14&  1.16$\pm$0.23& 0.84$\pm$0.14&\\
$14_{1,14}-13_{1,13}$&47271.282$\pm$0.010& 0.83$\pm$0.12&  0.70$\pm$0.11& 1.10$\pm$0.18&\\
$14_{0,14}-13_{0,13}$&47971.107$\pm$0.010& 0.73$\pm$0.10&  0.65$\pm$0.11& 1.07$\pm$0.16&\\
$14_{1,13}-13_{1,12}$&49042.693$\pm$0.010& 0.73$\pm$0.10&  0.65$\pm$0.11& 1.07$\pm$0.16&\\
\hline
\end{tabular}
\tablefoot{
\tablefoottext{a}{Observed frequencies for the detected transitions of NCCHCS in TMC-1.}\\
\tablefoottext{b}{Integrated line intensity in mK\,km\,s$^{-1}$.}\\
\tablefoottext{c}{Line width at half intensity using a Gaussian fit in the line profile
(in km~s$^{-1}$).}\\
\tablefoottext{d}{Antenna temperature (in mK).}\\
\tablefoottext{A}{Only data from the observations with a frequency switching of 10 MHz.}\\
\tablefoottext{B}{The line is fully blended with the rotational transition
$8_{17}-7_{16}$ of H$_2$CCCS.}\\
\tablefoottext{C}{Only data from the observations with a frequency switching of  8 MHz.}
}\\
\end{table*}

\end{appendix}
\end{document}